\documentclass[twocolumn,twocolappendix]{aastex631}

\usepackage[utf8]{inputenc}

\usepackage{newtxtext,newtxmath}

\usepackage[T1]{fontenc}
\usepackage{ae,aecompl}
\usepackage[mathscr]{euscript}

\usepackage{float}
\usepackage{graphicx}
\usepackage{subfig}
\usepackage{amsmath}
\graphicspath{{./Images/}}
\usepackage{float}
\usepackage{xcolor}
\usepackage{subfig}

\newcommand{\msun}{$\rm M_{\odot}$}

\newcommand{\beq}{\begin{equation}}
\newcommand{\eeq}{\end{equation}}
\newcommand{\beqn}{\begin{eqnarray}}
\newcommand{\eeqn}{\end{eqnarray}}

\defcitealias{SG}{SG}
\defcitealias{TQM}{TQM}
\defcitealias{agnpack}{\texttt{pAGN}}
\defcitealias{pAGN}{Gangardt et al.~(2024)}

\usepackage{color}
\definecolor{cerulean}{rgb}{0.0, 0.48, 0.65}

\definecolor{navy}{rgb}{0.2, 0.0, 1.0}

\definecolor{jungle}{rgb}{0.0, 0.5, 0.0}

\makeatletter
\def\@to{to}
\makeatother

\shorttitle{AGN Stars}
\shortauthors{Fabj et al.}

\begin{document}
\title{Mapping the Outcomes of Stellar Evolution in the Disks of Active Galactic Nuclei}

\correspondingauthor{Gaia Fabj}
\email{gaia.fabj@nbi.ku.dk}

\author[0000-0002-8875-3116]{Gaia Fabj}
\affil{Niels Bohr International Academy, The Niels Bohr Institute, Blegdamsvej 17, 2100 Copenhagen, Denmark}

\author[0000-0001-6157-6722]{Alexander J.~Dittmann}
\affil{Department of Astronomy and Joint Space-Science Institute, University of Maryland, College Park, MD 20742-2421, USA}
\affil{Theoretical Division, Los Alamos National Laboratory, Los Alamos, NM 87545, USA}

\author[0000-0002-8171-8596]{Matteo Cantiello}
\affiliation{Center for Computational Astrophysics, Flatiron Institute, New York, NY 10010, USA}
\affiliation{Department of Astrophysical Sciences, Princeton University, Princeton, NJ 08544, USA}

\author[0000-0002-3635-5677]{Rosalba Perna}
\affiliation{Department of Physics and Astronomy, Stony Brook University, Stony Brook, NY 11794-3800, USA}

\author[0000-0003-0607-8741]{Johan Samsing}
\affiliation{Niels Bohr International Academy, The Niels Bohr Institute, Blegdamsvej 17, 2100 Copenhagen, Denmark}

\begin{abstract}

The disks of Active Galactic Nuclei (AGNs) are expected to be populated by numerous stars, either formed in the outer regions of the disk via gravitational instability, or captured from the nearby nuclear star cluster. Regardless of their formation mechanism, these stars experience altered evolutionary paths, mostly shaped by the accretion of dense disk material.
In this study, through the comparison of different timescales, we chart the evolutionary outcomes of these AGN stars as a function of disk radius and across a range of supermassive black hole (SMBH) masses, spanning from $10^6$ to $10^9 \rm M_\odot$, for two popular AGN disk models. 
We find that, in the outer regions of the disk, stars evolve similarly to those in the interstellar medium, but in the inner and denser regions accretion quickly turns low-mass stars into massive stars, and their fate depends on just how quickly they accrete. If accretion occurs at a faster rate than nuclear burning, they can reach a quasi-steady `immortal' state. 
If stars accrete faster than they can thermally adjust, runaway accretion occurs, potentially preventing a quasi-steady state and altering the disk structure. 
During the AGN lifetime,
in the regions of the disk that produce massive stars, supernovae (SNe) and 
Gamma-Ray Bursts (GRBs) may occur within the disk over a wide range of optical depths and ambient densities.
Subsequently, in the final phase of the AGN, as the disk becomes depleted, formerly immortal stars will be unable to replenish their fuel, leading to 
additional SNe and GRBs.

\end{abstract}


\section{Introduction}

Active Galactic Nuclei (AGNs) are powered by large disks of gas accreting onto a central supermassive  black hole (SMBH). While historically the focus has largely been on the properties of the accreting gas \citep{1969Natur.223..690L,1973A&A....24..337S}, recently, interest has grown in the study of accretion disks as the hosts of stars and the compact 
objects that they leave behind. 

Stars are believed to exist in AGN disks as a result of two mechanisms. The outer regions of these disks, prone to become gravitationally unstable, may likely form stars \citep[e.g.,][]{1980SvAL....6..357K,2003MNRAS.339..937G,Dittmann2020,2022arXiv220510382D}. Additionally, stars from the nuclear star cluster  with misaligned orbits with respect to the disk get captured over time by gaseous torques \citep[e.g.,][]{1993ApJ...409..592A,2004ApJ...602..388T,Fabj20,Nasim23, Wang2024}.
The presence of stars in AGN disks is closely intertwined with the presence of the compact objects that they leave behind, and that is white dwarfs, neutron stars and black holes, with the last two of particular interest in connection to LIGO/Virgo observations of gravitational waves when they merge. In fact, much recent work on objects in AGN disks has been motivated by unexpected discoveries by LIGO/VIRGO, such as BHs in the lower mass gap \citep[e.g.,][]{Abbott2020low,Yang2020,Tagawa2020} as well as in the upper mass gap \citep[e.g.,][]{Abbott2020high,GerosaFishbach21,Ford22}, and an asymmetry in the BH spin distribution \citep[e.g.,][]{Callister2021,McKernan2021,Wang2021a}.

Regardless of the mechanisms by which stars end up in an AGN disk, their evolution can be profoundly affected by this exotic environment. The very high densities can result in stars accreting large fractions of their initial masses during their lifetimes, potentially reaching a quasi steady-state with no chemical evolution. These stars, for which accretion of fresh gas compensates for the fuel used to power them, can live a very long time, in principle as long as the AGN disk itself, and are dubbed "immortal" \citep{Paper1,Paper2,paper3,Paper4}. 
Additionally, because stars gain angular momentum as they accrete, many will reach the end state of their evolution endowed with a large enough angular momentum in their core and in the fallback envelope material to create ripe conditions to yield gamma-ray bursts (GRBs), in addition to supernovae, when they explode \citep{Paper2}.

Considering the potentially significant influence of the AGN disk environment on star evolution, it follows that stars initially identical but situated in varied disk locations may undergo distinct evolutionary paths.
Charting these paths as a function of the AGN disk location and disk mass is thus an important step towards assessing the frequency of compact-object related events, 
in AGN disks, such as GRBs \citep[e.g.,][]{Perna2021a,Zhu2021GRB,Lazzati2022}, micro-tidal disruption events \citep{Perets2016,Yang2022,Prasad2024}, accretion-induced collapse of white dwarfs \citep{Zhu2021WD} and neutron stars \citep{Perna2021b}, binary black hole mergers \citep[e.g.,][]{Bartos17,McKernan2020,Samsing2022,Tagawa2023m,2024arXiv240216948F}, accreting black holes \citep{Tagawa2023s}, Fast radio Bursts \citep{Zhao2024},
among others. Such transients may contribute to the bright flares observed in many AGN, or other forms of AGN variability (see e.g. \citealt{Graham2017,Wang2021AGNstars,Liu2024}).
Conversely, such a mapping of stellar outcomes would also allow to better assess the influence of the stars themselves on  AGN disks \citep{Jermyn2022}.

Our paper is organized as follows:
In Sec.~2 we describe the disk models we adopt for our study. The results of the star mapping within the disk are presented in Sec.~3. We discuss the implications of our findings in Sec.~4 and summarize and conclude in Sec.~5.

\section{Disk Models}\label{sec:models}
We test two disk models: \citet{SG} (\citetalias{SG} in the following) and \citet{TQM}(\citetalias{TQM} in the following), which we calculate using the \citetalias{agnpack} package from \citetalias{pAGN}. 
These two models attempt to extend the standard accretion disk model of \citet{1973A&A....24..337S} to $\gtrsim {\rm pc}$ scales in SMBH accretion disks. Specifically, both \citetalias{SG} and \citetalias{TQM} attempt to ameliorate the shortcoming of \citet{1973A&A....24..337S} first pointed out in \citet{1980SvAL....6..357K}: that the disk models become gravitationally unstable at large radii and might collapse to form stars. Both \citetalias{SG} and \citetalias{TQM} solve this problem by invoking additional heating in the disk, to the extent necessary to maintain gravitational stability ($Q\geq1$, \citealt{1964ApJ...139.1217T}). Additionally, both the \citetalias{SG} and \citetalias{TQM} models are typically computed using tabulated opacity tables, mapping between gas and dust opacities at high and low temperatures, whereas \citet{1973A&A....24..337S} focused on temperatures where analytical opacity approximations were applicable.

In standard viscous accretion disk models, heat generated by viscous dissipation is radiated away locally, directly relating the emergent flux and effective temperature of the disk to its accretion rate and angular velocity profile \citep[e.g.,][]{1974MNRAS.168..603L}, $T_{\rm eff}^4\propto \dot{M}\Omega^2$ away from the disk's center. \citetalias{SG} does away with this equation in the outer regions of the disk, instead enforcing
\begin{equation}\label{eq:rhoStable}
\rho = \frac{\Omega^2}{2\pi G Q_{\rm crit}},
\end{equation}
where $\Omega$ is the angular velocity of the AGN disk at a given location and $Q_{\rm crit}\sim1$, the critical value of $Q$ for gravitational stability, is a free parameter. The models in \citetalias{SG} included extra heating to the extent required to stabilize the disk. In this work, we have computed these models using  opacity tables from \citet{2003A&A...410..611S} at temperatures below $10^4$~K, and
from \citet{2005MNRAS.360..458B} at higher temperatures.

On the other hand, the disk models presented in \citetalias{TQM} explicitly account for the mass consumed by star formation in the disk, where feedback is accounted for by setting $Q=1$, with disk structure primarily shaped by radial variations in accretion rate driven by star formation.
The efficiency for that matter to convert to energy and return it to the disk as radiative feedback ($\epsilon$), and the additional radiation pressure support in the disk thanks to this feedback. Crucially, this treatment causes the gas accretion rate through the disk to change as a function of radius as it forms stars. The models presented in 
\citetalias{TQM}
also include a non-Keplerian angular velocity profile at large distances, based on a fiducial velocity dispersion. Similarly to the \citetalias{SG} case, here as in \citetalias{pAGN}, we use the opacity blend from \citet{2003A&A...410..611S} at low temperatures (below $\sim 10^4$~K) and \citet{2005MNRAS.360..458B} at higher temperatures.\footnote{This differs from the models calculated in 
\citetalias{TQM},
which used \citet{2003A&A...410..611S} at low temperatures but assumed electron scattering opacity otherwise. } We have also assumed $\epsilon=10^{-3}$, 
using the standard assumptions of the \citetalias{TQM} paper (energy generation by fusion and a Salpeter initial mass function). While \citetalias{SG} uses a \citet{1973A&A....24..337S} $\alpha$ viscosity prescription, \citetalias{TQM} 
uses a constant radial Mach number formulation instead, where we set the value of advection efficiency $m_t = 0.2$ as in the original \citetalias{TQM} model. As $m_t$ (ratio between the radial velocity of the advection and local sound speed) is set to be <1, advection occurs at a sufficiently slow pace, ensuring it does not significantly influence the accretion process onto embedded stars.
Furthermore, for our main results we set $\alpha=0.01$, and disk Eddington ratio (ratio between the emitted luminosity and the Eddington luminosity) $l_{\rm E}=0.5$, but also explore how setting $\alpha=0.1$ and $l_{\rm E}=0.05$ change our results in Appendix~\ref{appendix:alpha}. 
For \citetalias{TQM} we set the outer accretion rate to $\dot{M}_{\rm out} = 320 M_{\odot}/\rm yr \times (M_{\rm SMBH}/10^8 M_{\odot})^2$, in order to achieve roughly constant Eddington ratios for the accretion rate onto the SMBH.

\begin{figure}
    \centering
    \includegraphics[width=1.\columnwidth]{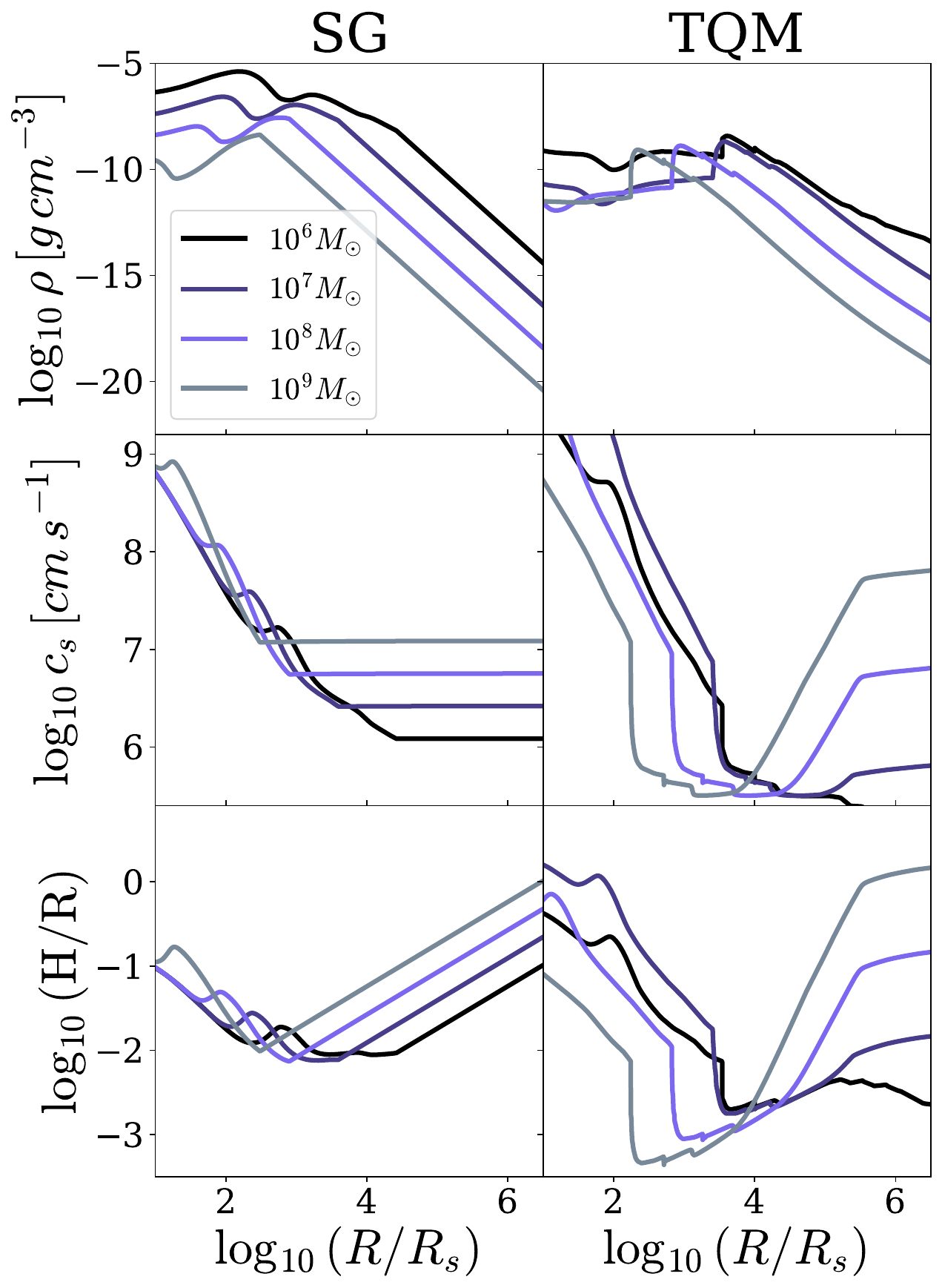}
    \caption{Density (top), sound speed (middle) and scale height (bottom) radial profiles for the  
    \citetalias{SG} (left) and \citetalias{TQM} (right) disk models, each  for four values of the SMBH mass in the $10^6\, \rm M_\odot$ - $10^9 \, \rm M_{\odot}$ mass range.
    }\label{fig:diskProfiles}
\end{figure}

A few example disk models are presented in Figure \ref{fig:diskProfiles}. 
From top to bottom, we display the radial profiles of the central density, the sound speed, and the scale height (in units of the 
Schwarzschild radius $R_s=2GM_{\rm SMBH}/c^2$) for four representative SMBH masses ($M_{\rm SMBH}$) between $10^6-10^9M_\odot$. The left and right panel contrast the same profiles for the \citetalias{SG} and the \citetalias{TQM} models, respectively.

Although our current study is restricted to the disk models presented in \citetalias{SG} and \citetalias{TQM}, which should provide a solid qualitative picture of the general regions where different outcomes of stellar evolution should occur, a number of other disk models exist. For example, within the framework of \citetalias{TQM}, \citet{Dittmann2020} pointed out that accretion onto embedded stellar-mass black holes would provide much more efficient feedback than would star formation, helping stabilize accretion disks with less mass depletion. More recently, \citet{2022ApJ...928..191G} and \citet{EpseteinMartin2024} have developed time-dependent disk models. Others have developed disk models invoking magnetic support, which might also help prevent star formation \citep[e.g.,][]{2007MNRAS.375.1070B,2024OJAp....7E..20H}. 

\begin{figure*}
    \centering
    \includegraphics[width=\linewidth]{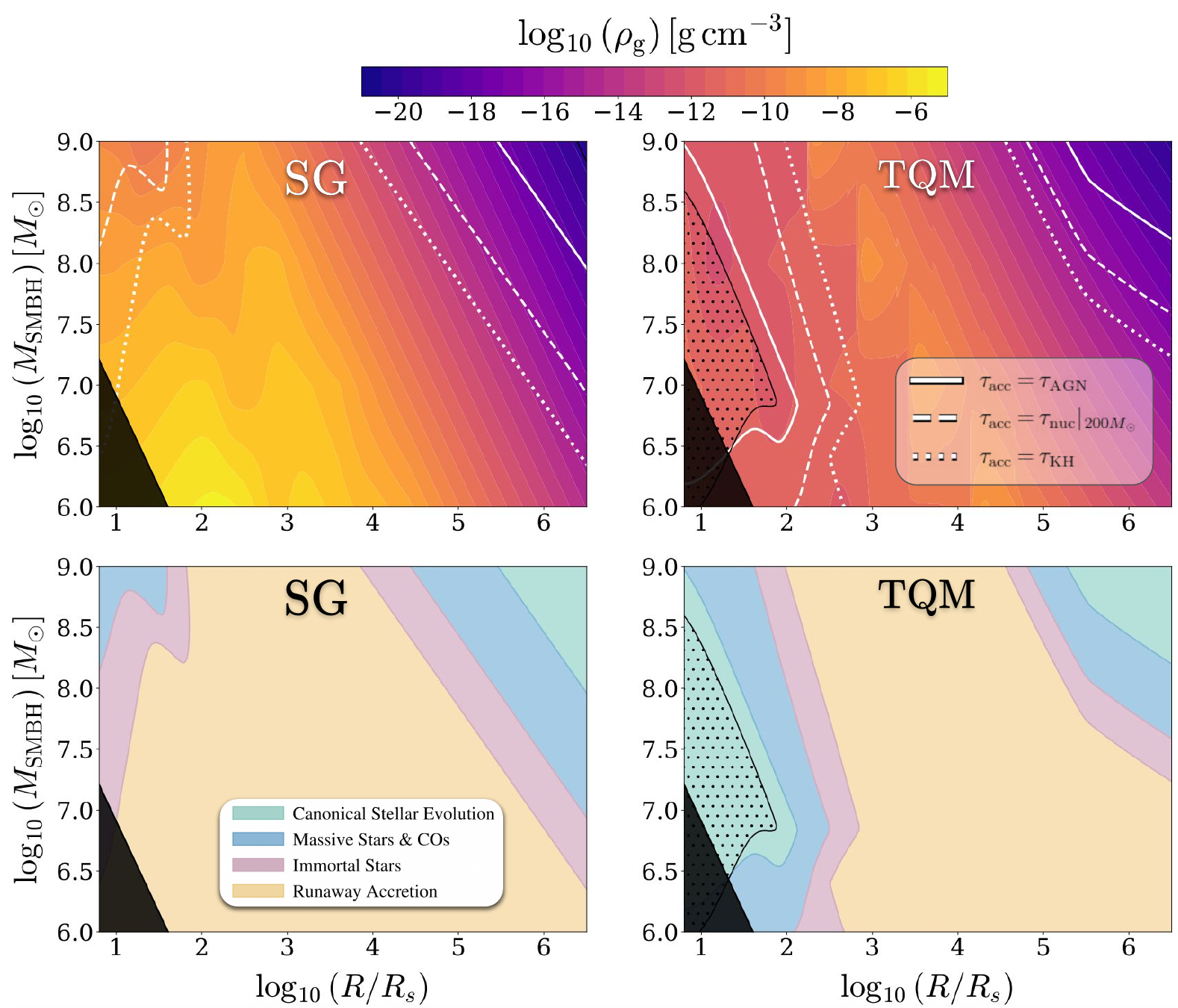}
    \caption{Upper panels: density contour maps for \citetalias{SG} (left) and \citetalias{TQM} (right) for different SMBH masses in the range of $10^6-10^9 \, \rm M_{\odot}$. The over-plotted white lines indicate the conditions based on comparing $\tau_{\rm acc}$ to the other relevant timescales ($\tau_{\rm AGN}, \, \tau_{\rm nuc}|_{200 M_\odot}, \, \tau_{\rm KH} $) as explained in Sec.~\ref{sec:map}. The lower panels display different stellar evolution outcomes based on the timescales comparison: canonical stellar evolution(green), massive stars \& COs (blue), immortal stars (pink), runaway accretion(yellow). The dotted contour region indicates where the models become unphysical ($c_s \sim c$).
    The dark region in the bottom corner of all four panels indicates the tidal limit for a star of radius of 10~$R_\odot$  and mass of 200 M$_\odot$. As the disk density increases, the mass accretion onto the star is more significant and stellar evolution is significantly affected.}
    \label{fig:rhomap}
\end{figure*}

\begin{figure*}
    \centering
    \includegraphics[width=1.\linewidth]{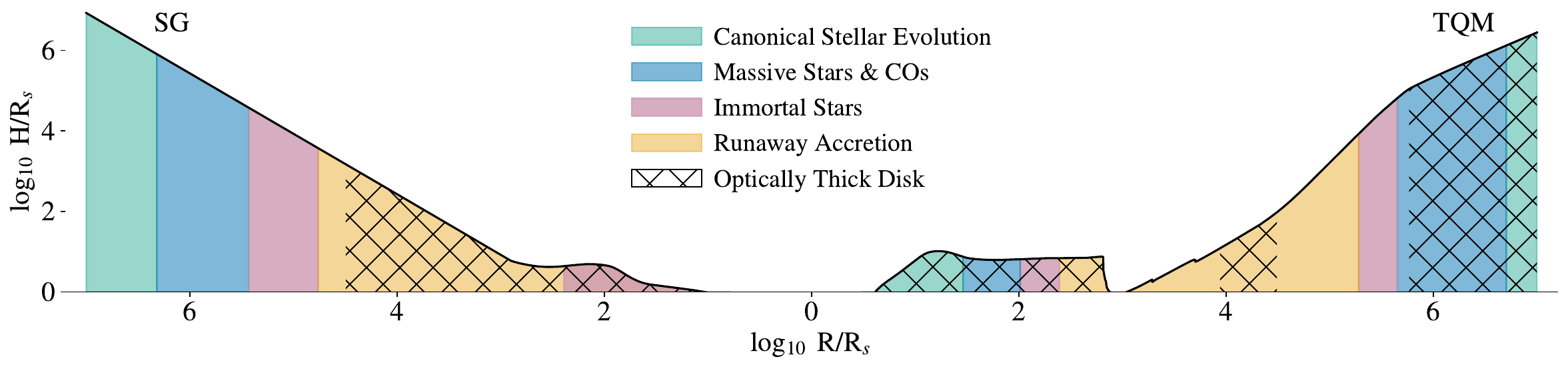}
    \caption{Different outcomes of stellar evolution in the disk using the \citetalias{SG} (left) and \citetalias{TQM} (right) modeled for a $10^8 \, M_{\odot}$ SMBH. The x-axis shows distance from the central SMBH in terms of $R_s$, while the y-axis shows the disk scale height H in the same units.
    As in the lower panels of Fig.~\ref{fig:rhomap}, colors indicate the different evolutionary regions. Hatched zones correspond to optically thick disk regions ($\tau > 1 $). Runaway accretion occurs in a large portion of the disk for both models, and in the case of \citetalias{TQM} stellar evolution goes back to its canonical form in the innermost parts of the disk given  the reduced accretion
    rate due to the combination of tidal effects with lower gas density and higher sound speed.
    }
    \label{fig:agn_chart}
\end{figure*}

\section{Mapping the evolutionary regions of AGN stars} 
Previous investigations have used suites of stellar evolution simulations to study in detail the dependence of stellar evolution on AGN disk properties such as density, sound speed, and angular velocity, and to derive power-law expressions relating evolutionary outcomes to these disk parameters. However, the conditions within AGN disks cover a range of conditions far more broad than those covered by stellar evolution simulations. The different outcomes of stellar evolution in AGN disks can also be quantified by comparing the relevant timescales, particularly those for stellar accretion, nuclear burning, and thermal adjustment \citep{Paper4}. Below, we use these timescales to generalize previous estimates so that we can infer the outcomes of stellar evolution in AGN disks over their full range of conditions.

\subsection{Comparison Between Different Timescales}\label{sec:map}

In order to quantify the effect of the AGN disk on canonical stellar evolution, we first compute the stellar accretion timescale $\tau_{\rm acc}$ based on the disk local properties. 
The accretion rate depends on whether the region around the star is approximately spherically symmetric, or affected by tidal forces from the SMBH. To establish which regime is dominating, we compute the Bondi radius $R_B$ and the Hill radius $R_H$, where the former is defined as:
\begin{equation}
R_B \equiv \frac{2GM_*}{c_s^2},
\end{equation} 
where $M_*$ is the mass of the star and $c_s$ is the disk local sound speed. 
We define the Hill radius as:
\begin{equation}
R_H \equiv \left(\frac{GM_{*}}{3\Omega^2}\right)^{1/3},
\end{equation}
where $\Omega$ is the star orbital frequency. 
If $R_H < R_B$ we are in the tidally-limited regime, while if $R_B< R_H$ we are in the Bondi accretion-dominated regime. We compute the mass accretion rate using the following expression:
\begin{equation}\label{eq:mdot}
\dot{M} = \frac{\pi}{f}\rho c_s\left(\min\left\{R_B,R_H \right\}\right)^2,
\end{equation}
where $\rho$ is the disk density and $f\sim4$ is a luminosity-dependent factor that accounts for the decrease of the accretion rate onto the star as it approaches the Eddington luminosity (see \citealt{Paper1} for details).
 The accretion timescale at any location in the disk is thus given by
 $\tau_{\rm acc}\equiv M_*/\dot{M}$, which can be rewritten as 
 
\begin{equation}\label{eq:tau_M}
\tau_{\rm acc} \equiv \frac{fM_*}{\pi}\frac{1}{\rho}\frac{1}{c_s}\left(\min\left\{R_B,R_H \right\}\right)^{-2}.
\end{equation}
The relevant $M_{*}$ depends on the boundary between different stellar evolution types in the disk.

From \citet{paper3}, the typical masses for immortal stars are in the range of 200 to 1000 $M_{\odot}$. 
We therefore adopt 1 $M_{\odot}$ for the boundary between canonical stellar evolution and massive stars, and 200 $M_{\odot} $ to draw the boundary for immortal and runaway stars \footnote{Changes in the AGN disk composition would  lead to a factor of order of unity changes in the chosen values \citep{Paper4}.}.
The minimum radius in Eqs.~\ref{eq:mdot} \& \ref{eq:tau_M}   stands to indicate that accretion occurs at the slower rate, and therefore at the longer timescale.

In order to map the outcome of stellar evolution in the different regions of the disk, we draw a comparison between $\tau_{\rm acc}$ and the other relevant timescales: AGN-disk lifetime, nuclear-burning, and Kelvin-Helmholtz timescales.
To determine if the star is affected by the surrounding gas within the period of the galaxy's active phase, we compare the star accretion timescale to the AGN disk lifetime $\tau_{\rm AGN}$, which we set to a fiducial value of $10^7$ years. 

We therefore establish the criterion that when
\begin{equation}\label{eq:crit}
\tau_{\rm acc} > \tau_{\rm AGN},
\end{equation}
the star undergoes canonical stellar evolution. 
As we get to parts of the disk with higher densities, the star undergoes significant accretion, becomes massive, and can potentially evolve into a compact object. 
The star does not however necessarily become immortal, since in order to enter the immortal phase, the star has to satisfy additional conditions. As shown in \citet{Paper4}, a star can be considered immortal if it accretes at a faster or at the same rate than it can burn its entire reservoir of hydrogen. We thus additionally compare $\tau_{\rm acc}$ to the nuclear burning timescale for a high-mass star.  
The nuclear-burning timescale for a 200 $M_{\odot}$ star can be approximated to:
\begin{equation}\label{eq:taunuc}
\tau_{\rm nuc}|_{200M_\odot} \approx 5.5\times10^{10}\left(\frac{M_*}{M_\odot}\right)\left(\frac{L_*}{L_\odot}\right)^{-1} {\rm yr}\approx 1.7\times10^6 {\rm yr},
\end{equation}
being $L_*$ the stellar luminosity, where the approximation assumes radiation at $L_*\approx L_{\rm Edd}\approx 3.2\times10^4(M/M_\odot)$. The pre-factor of $5.5$ was chosen to reproduce the results in Section 4.3 of \citet{paper3} under the assumptions of $L_*=L_{\rm edd}$, $f=4$, and $M_*=200M_\odot$. \footnote{A star made of pure hydrogen, able to consume all of its fuel, would have a nuclear burning timescale of $\tau_{\rm nuc}\approx10^{11}\left(\frac{M_*}{M_\odot}\right)\left(\frac{L_*}{L_\odot}\right)^{-1}$ years. Since we are considering fully convective stars with $X\sim0.7$, a pre-factor of $5.5$ is appropriate, although pre-factors for the nuclear burning timescales for lower-mass stars are closer to $\sim 1$.}

If the accretion timescale is shorter than the AGN lifetime but longer than the nuclear burning timescale, that is, 
\begin{equation}\label{eq:crit2}
  \tau_{\rm nuc}|_{200M_\odot}\leq \tau_{\rm acc} \leq \tau_{\rm disk}, 
\end{equation}
we can have the formation of massive but "mortal" stars, which can potentially end their lives as compact objects. 

In order to map the regions where immortal stars are formed, we draw the boundaries where the accretion timescale is shorter than the nuclear burning timescale and longer than the stellar inner thermal adjustment timescale. The latter is set by the Kelvin-Helmholtz timescale $\tau_{\rm KH}$, which is approximately $3\times10^4$ years for massive stars \citep{Bond1984}. 
This leads to setting the following criterion for immortal stars: 

\begin{equation}\label{eq:crit3}
\tau_{\rm KH}<\tau_{\rm acc}\leq \tau_{\rm nuc}|_{200 M_\odot},
\end{equation}
where, as long as $\tau_{\rm acc}>\tau_{\rm KH}$, the star remains stable, as the stellar luminosity is able to respond to changes in the stellar mass while maintaining  equilibrium.
However, if $\tau_{\rm acc}$ becomes shorter than the stellar thermal adjustment time, the stellar luminosity cannot adjust the mass loss rate to compensate accretion, and the star undergoes runaway accretion. 
Therefore, we use the following condition 
\begin{equation}\label{eq:crit4}
    \tau_{\rm acc} \le \tau_{\rm KH},
\end{equation}
to identify where stars undergo runaway accretion. 

The upper panels of Fig.~\ref{fig:rhomap} show the density contour plot of \citetalias{SG} (left panels) and \citetalias{TQM} (right panels) for different SMBH masses in the range of $10^6-10^9 \, \rm M_{\odot}$. 
The white lines indicate the boundaries that separate the different regions of stellar populations based on the balances between the different timescales. 
The solid lines indicate where $\tau_{\rm acc}=\tau_{\rm AGN}$,
the dashed lines mark $\tau_{\rm acc}=\tau_{\rm nuc}|_{200M_\odot}$, while the dotted ones represent where $\tau_{\rm acc} = \tau_{\rm KH}$. Therefore, 
the regions in between the lines mark the different types of stellar evolution and where they can occur based on the disk properties. 
The green region in the lower panels of Fig.~\ref{fig:rhomap} indicates where 
Eq.~\ref{eq:crit} is true and hence where in the disk stars undergo canonical stellar evolution. 
Hence  stars in this region are not significantly affected within the disk lifetime and undergo regular stellar evolution. 
Massive stars and COs are found in the blue area, in regions of the disk where Eq.~\ref{eq:crit2} is true, while the condition of immortal stars satisfied by Eq.~ \ref{eq:crit3} is indicated by the pink region. 
Lastly, the yellow area represents where in the disk otherwise-immortal stars undergo runaway accretion and become unstable, under the condition of Eq.~\ref{eq:crit4}.
The dotted black region in both panels of the \citetalias{TQM} model shows where the model becomes unphysical as the gas sound speed $c_s$ approaches the speed of light $c$. 
The area marks where in the disk $c_s/c>0.1$.
In all four panels, the shaded black region represents the tidal limit for a star of radius 10 $R_{\odot}$ and 200 $M_{\odot}$. 
In the case of \citetalias{SG}, as shown in Fig.~\ref{fig:rhomap}, the dominant process at low SMBH masses is runaway accretion, while immortal stars 
can form for masses greater than $10^7 \, M_{\odot}$ in both inner and outer parts of the disk. 
Canonical stellar evolution can be present for $M_{\rm SMBH} > 10^7$ $M_{\odot}$ and it mostly takes place in the outer parts of the disk, with the region increasing 
in size with increasing $M_{\rm SMBH}$. 
The reason for the \citetalias{SG} model to have such a restricted region of canonical stellar evolution in the inner parts of the disk is mainly due to our choice of disk parameters, such as $\alpha$ and $l_{\rm E}$, which we vary and investigate the subsequent impact in Appendix~\ref{appendix:alpha}.
In \citetalias{TQM} disks, runaway accretion is more constrained and canonical stellar evolution can occur in both inner (for $M_{\rm SMBH} > 10^6$ $M_{\odot}$) and outer regions (for $M_{\rm SMBH} \ge 10^8$ $M_{\odot}$).
The behavior is due to the fact that in the \citetalias{TQM} inner region  the local disk sound-speed is significantly higher, corresponding to less efficient accretion. 

We analyze more in detail the different locations of the evolutionary types by plotting the stellar evolution maps for 
a single SMBH mass of $10^8 \, M_{\odot}$ in Fig.~\ref{fig:agn_chart}. 
In order to better visualize the relationship between the different accretion regions and disk structure, we plot the disk scale-height $H$ in terms of $R_s$, as a function of distance from the central object. We also show the parts of the disk with high optical depth ($\tau>1$) and therefore where the objects are obscured. For a $10^8 \, M_{\odot}$, the inner parts of the \citetalias{SG} model ($R<10^5 \, R_s$) are obscured. 
In addition, in the case of a 
$10^8 \, M_{\odot}$ SMBH, \citetalias{SG} has a larger region than the \citetalias{TQM}  disk where immortal stars can exist, while for both models the evolution into "mortal" massive stars \& COs occurs 
in a more restricted portion of the disk compared to the other evolutionary types. 
Fig.~\ref{fig:agn_chart} also shows that for \citetalias{SG} the disk is optically thick up to $\sim 5\times 10^4 R_s$, and therefore obscured by dust, while in the case of \citetalias{TQM}, the outer parts of the disk become optically thick as well. 
We investigate in further detail the accretion disk optical depth in Sec. \ref{sec:tau}.

\subsection{Mapping the Optical Depth}\label{sec:tau}
\begin{figure*}
    \centering
    \includegraphics[width=\linewidth]{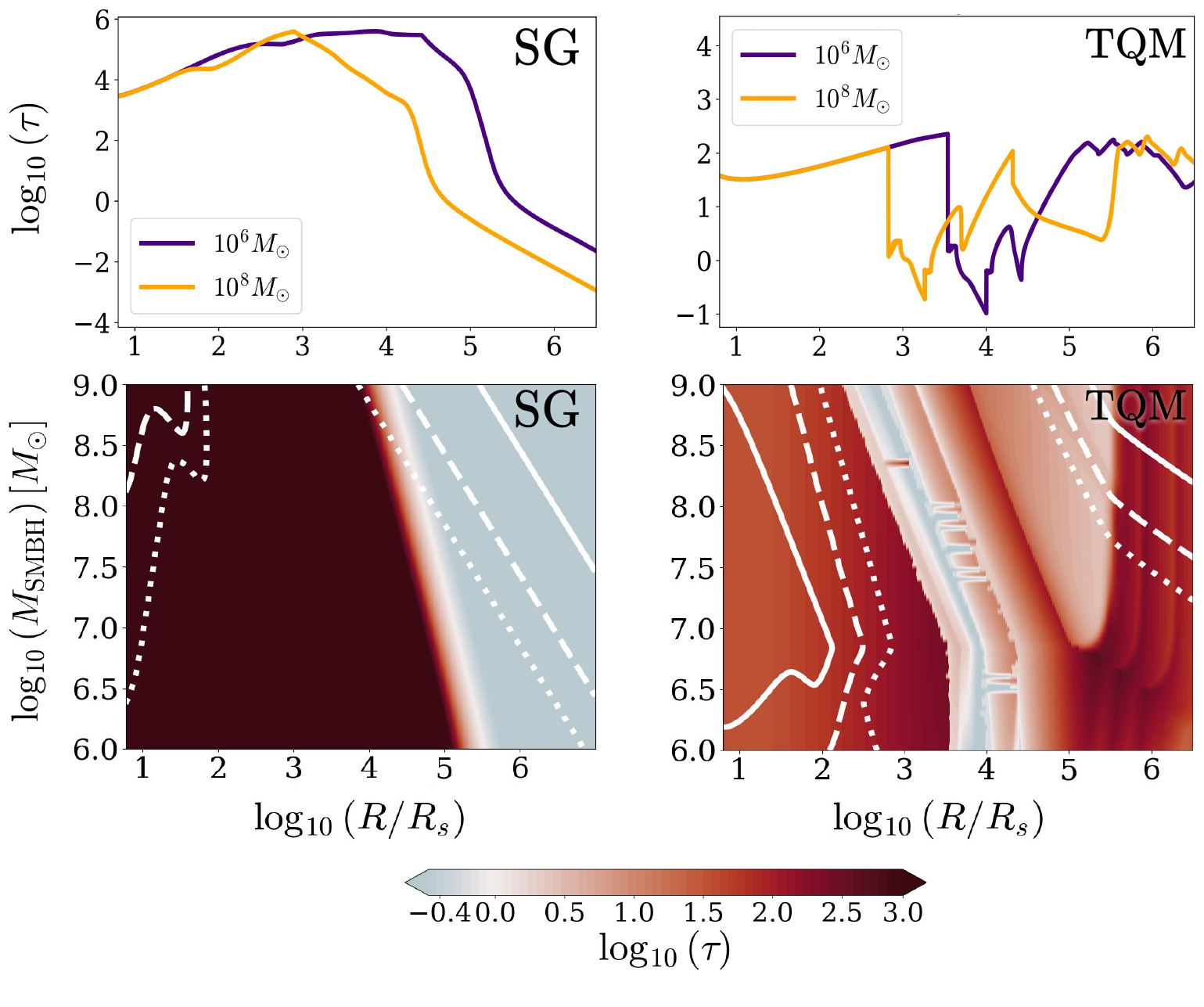}
    
    \caption{Optical depth $\tau = \kappa \rho H$ 
    for the \citetalias{SG} (left) and 
    \citetalias{TQM} models (right).
    In both cases, the opacity $\kappa(\rho,T)$ is built from the 
    tables of \citet{2003A&A...410..611S} for $T < 10^4$~K, and from the tables of \citet{2005MNRAS.360..458B} at higher temperatures. Given the same $\kappa$ prescription, the different behaviors of the optical depth in \citetalias{SG} and \citetalias{TQM} are due to their different temperature profiles (see $c_s$ profile in the mid-panel of Fig~\ref{fig:diskProfiles} for reference). 
    The top panels show the optical depth radial profile for two specific values of the SMBH mass, $10^6 \, M_{\odot}$ (purple) and $10^8 \, M_{\odot}$ (orange). 
    These diagrams provide intuition for the 2D plots in the bottom, where  the optical depth is illustrated for the full range of SMBH masses of this study. Overlaid are the lines demarcating the regions where the different types of evolutionary outcomes for AGN stars are expected.
    }
    \label{fig:taumap}
\end{figure*}

The following section aims to describe the different features of the AGN disk optical depth $\tau$ depending on the chosen disk model and to discuss the implications, as in which type of objects lie in the optically thin region and can therefore be considered "visible" and un-obscured by the surrounding gas.
The four panels in Fig.~\ref{fig:taumap} display the optical depth for \citetalias{SG} (left panels) and \citetalias{TQM} (right panels). 
The upper panels show $\tau$ as a function of distance from the SMBH for two different masses ($10^6 \, M_{\odot}$ in purple and $10^8 \, M_{\odot}$ in orange), while the lower panels show $\tau$ contour plots for different $M_{\rm SMBH}$ in the range of $10^6-10^9 \, M_{\odot}$. 
For both models, the optical depth at the midplane is evaluated through $\tau = \kappa \rho H $, where $\kappa$ is the disk opacity $\kappa(\rho,T)$, which is built from the tables of \citet{2003A&A...410..611S} for $T < 10^4$~K, and from the tables of \citet{2005MNRAS.360..458B} at higher temperatures. 
As seen in both the lower and upper panels of Fig.~\ref{fig:taumap}, the optical depth profiles of \citetalias{SG} and \citetalias{TQM} display different features. In the \citetalias{SG} model, the optically thin regions ($\tau \le 1$) are mainly concentrated in the outer parts of the disk ($10^5-10^6 \, R_s$). 
With respect to \citetalias{SG}, \citetalias{TQM} displays sharp features and a more irregular trend, and the regions of low optical depth are mainly concentrated between $10^3-10^4 \, R_s$. 
The sharp features of $\tau$ in \citetalias{TQM} are mainly due to the opacity dependence on temperature. At lower temperatures ($10^3-10^4$ K) dust sublimation occurs and opacity becomes negligible \citep{2003A&A...410..611S}. 
On the other hand, since \citetalias{SG} tends to have a hotter disk, opacity has a much stronger influence than dust sublimation. 
In the lower panels of Fig.~\ref{fig:taumap} 
we also show the boundary lines of the different stellar evolution outcomes as described in Sec.~\ref{sec:map} and represented in Fig.~\ref{fig:rhomap}. In the case of \citetalias{SG}, the optically thin region covers all the four different types of evolutionary regions: canonical (outermost region), massive stars, immortals and runaway accretion (innermost region). On the other hand, the \citetalias{TQM} opacity map shows that the disk is optically thin in the inner regions, where stellar runaway accretion dominates.
\section{Discussion}

In the following, we discuss the implications of our findings for the stars in each regions, and for the AGN disk.

\subsection{Runaway Accretion}\label{sec:discussRunaway}
Runaway accretion occurs when the a star accretes more quickly than stellar structure can adjust, making reaching a steady (or quasi-steady) state impossible.\footnote{In some previous work, particularly \citet{Paper1}, \citet{Paper2}, and \citet{paper3}, ``runaway accretion'' has been used to describe the regime we describe here as ``immortal stars.'' Here, we draw distinction between these equilibrium (``immortal'') and disequilibrium (``runaway'') regimes.} This realm of stellar evolution has proven difficult to simulate due to numerical challenges at the necessary accretion rates \citep[see, for example,][]{paper3}, so we instead draw from analytical estimates and studies of non-accreting stars.

In this particular limit, where the accretion timescale becomes shorter than the stellar thermal adjustment timescale, it is not necessarily a given that the star actually accretes all of the mass that enters its gravitational sphere of influence. If, as in protoplanetary accretion, the accreting gas can become pressure-supported, accretion will be limited by the cooling timescale of the gas \citep[e.g.,][]{2015MNRAS.447.3512O}. In this scenario, much of the gas that would have accreted onto the star instead dynamically outflows, often leading to patterns of meridional circulation. 
This might limit the accretion timescale to roughly the stellar Kelvin-Helmholtz timescale, potentially resulting in a marginally stable state. 

If stars are able to accrete unhindered, limited only by the available supply of gas, they will grow to such large masses that their interactions with the AGN disk become nonlinear, affecting its structure \citep{2004ApJ...608..108G}. 
In this limit, which occurs roughly when $M/M_\bullet \gtrsim (40\alpha)^{1/2}(H/R)^{5/2}$ \citep{1986ApJ...309..846L}, 
the star will carve out a large, deep gap in the disk substantially reducing the surface density of gas near the star, and thus its accretion rate \citep[e.g.,][]{2023MNRAS.525.2806C}. 
This would then truncate the stellar mass near $M\sim M_\bullet (40\alpha)^{1/2}(H/r)^{5/2}$. In  sufficiently thin disks ($H/R \sim 10^{-2}-10^{-3}$) gap opening count truncate stellar masses below $100 \, M_{\odot}$. Thicker disks may support stars in excess of $10^3-10^4 \, M_{\odot}$. 
Such stars may be general relativistically unstable \citep[e.g.,][]{1964ApJ...140..434T,1973ApJ...183..637B}. In particular, some of these stars may become unstable during hydrogen burning \citep{2023ApJ...949L..16N}, causing their ejecta to possess supersolar nitrogen abundances, offering a potential explanation for galaxies \citep[especially at high redshifts, e.g.,][]{2023MNRAS.523.3516C} and AGN \citep[e.g.][]{{2004AJ....128..561B},{2008ApJ...679..962J}} that display anomalously strong nitrogen lines. 

If these runaway stars leave behind commensurately massive remnants, intermediate mass black holes could be fairly common in AGN disks. The gravitational waves emitted by these objects will be detectable by the Laser Interferometer Space Antenna \citep[e.g.,][]{2017arXiv170200786A} and other future space-based interferometers such as Taiji and TianQin \citep[e.g.,][]{doi:10.1142/S0217751X2050075X,2016CQGra..33c5010L}, and may encode information about the AGN environment \citep[e.g.,][]{2021MNRAS.501.3540D,2023MNRAS.526.5612C}. The mergers of these intermediate mass black holes and their central SMBHs could also expedite the growth of SMBHs over cosmic time \citep[e.g.,][]{Dittmann2020}.

\subsection{Immortal Stars}
Immortal stars exist in a quasi-steady state where accreted mass balances are lost via winds. Since these stars are almost fully convective, radiation dominated, and likely rapidly rotating \citep{Paper1,Paper2}, H-rich material accreted from the disk is transported very quickly into the stellar core, constantly replenishing nuclear fuel.
It is known that stars above 100 $M_{\odot}$ might undergo instabilities that lead to an enhanced mass loss \citep{2016MNRAS.456..525G}. However, the outer structure of the immortal star is mainly dominated by the accretion stream that leads to a shocks on the surface, which might stabilize the stars.
Since energy generation in immortal stars is completely dominated by hydrogen fusion via the CNO cycle, the net effect of this process is to convert hydrogen from the disk into helium, at a rate dependent on the stellar luminosity \citep[e.g.,][]{Paper1, paper3}. For stars radiating at about the Eddington luminosity, this results in each star releasing helium into the disk at a rate of $\approx 10^{-4}\,{\rm M_\odot\,yr^{-1}}$; given a sufficient population of immortal stars, this might cause the AGN disk to become enriched with helium over time \citep{Jermyn2022}. If such enrichment occurs, stars would become more enriched in helium, thus more luminous at a given mass, and higher accretion rates would be required to sustain immortal stars \citep{Paper4}, slightly shifting inward the regions demarcated in Figures \ref{fig:rhomap} and \ref{fig:taumap}. However, given the fairly small range of radii in a given disk that can support immortal stars, such enrichment may not be common, instead limited to only disks with fairly large populations of stars in their outer regions. 

Immortal stars can remain as such as long as they reside within the AGN disk. However, as the disk is consumed during an AGN cycle, they evolve towards mortality as their source of fresh fuel disappears. Note that, as the disk dissipates, immortal stars may return mass fast enough to extend its lifetime by a factor of several, possibly driving powerful outflows from the disk \citep{Jermyn2022}. Since immortal stars have very high masses, once they stop accreting from the AGN disk they evolve towards a core collapse and a supernova (SN) explosion, where the observational signature will be only minimally influenced by the type of supernova or the ejecta mass since driven by optically thick material \citep{2021MNRAS.507..156G}.
Additionally, due to the high angular momentum acquired from accretion in a medium with a velocity shear \citep{Paper2}, the SN is likely to be accompanied also by a GRB. Therefore, the end of the AGN phase will be characterized by a burst of SNe/GRBs. 

Models show that both immortal and massive stars in AGN disks leave behind stellar mass black holes with typical mass around 10~\msun \citep{Paper1}. While this result is likely dependent on the uncertain physics of mass loss, the hint of a mass excess around this value in the underlying BH mass function inferred by LIGO-Virgo \citep{Abbott21,Abbott:2023} could be the signature of the population of BH produced by AGN stars \citep{Jermyn2022}. 

If a given AGN disk is particularly crowded with embedded objects, collisions between immortal stars or between immortal stars and compact objects may be fairly common. Given that AGN disks may capture or form as many as thousands of objects over their lifetimes \citep[e.g.][]{EpseteinMartin2024}, collisions might occur relatively frequently. 
The outcome of a collision between two immortal stars is highly uncertain, but since immortal stars are fully mixed and limited by mass loss at their Eddington luminosity, it is plausible that such collisions do not greatly change the nature of the immortal star population. The transient associated with such a massive stellar merger might be difficult to observe due to the optical depth of the dense surrounding environment.

If an immortal star and compact remnant collide, a quasi-star or Thorne-$\dot{\rm Z}$ytkow object may be formed, with the compact object acting as a stellar core \citep{1977ApJ...212..832T,2008MNRAS.387.1649B}.

\subsection{Massive Stars}

If the accretion rate is high 
enough to turn stars into massive stars, but the nuclear burning timescale is still shorter than the accretion timescale, we expect the formation of massive stars that can evolve and end their lives with stellar explosions, leaving behind compact remnants.

As discussed in Section~1, \citet{Paper2} demonstrated that the angular momentum contributed by accreting AGN gas to the stars makes it likely for massive AGN stars to produce a GRB together with a supernova. The potential to observe GRBs originating from AGN disks, along with the unique characteristics of their extreme host environments, is of considerable interest as such observations would offer direct insight into the stars within these disks.

\citet{Lazzati2022} showed that, for GRBs in high-density media, there exists a critical density,  $n_{\rm crit}\sim 10^6$~cm$^{-3}$
for average GRB parameters such as energy and Lorentz factor. Above this critical density, the external shock forms before the internal shock, leading to a rather smooth profile, devoid of the typical high variability of standard GRBs, and 
exhibiting a hard-to-soft spectral evolution. The duration is also generally longer than it would be in a standard galactic environment. Interestingly, a GRB with such characteristics was identified in the source GRB 191019A \citep{Lazzati2023} localized in the nucleus of a galaxy \citep{Levan2023}. 
In the disk models we studied (see Fig.~\ref{fig:rhomap}), the regions where massive stars are expected to form exhibit densities exceeding this critical value, and hence unusual prompt GRB transients are expected. 
Moreover, unlike standard GRB sources from galactic environments, GRBs from AGN disks tend to have their longer wavelength afterglow emission suppressed in comparison to the higher energy emission due to self-absorption \citep{Wang2022}.
This suppression would serve as another distinguishing feature of their emergence from the high density media of such environment.

Once produced in an AGN disk, the observability of GRBs will depend on the optical depth at their specific locations within the disk. As shown in Fig.~\ref{fig:taumap}, this optical depth is highly dependent on both the disk model and the SMBH mass.
 Notably, the \citetalias{SG} model is entirely optically thin in regions where massive star evolution is expected, whereas the \citetalias{TQM} model is predominantly optically thick.
It is important to note that the optical depth represented in Fig.~\ref{fig:taumap}
 reflects the opacity of the AGN disk before a transient event occurs. However, once GRB radiation begins to propagate through the disk, some of the X-ray photons may immediately ionize the disk. This ionization process shifts the opacity to be dominated by electron scattering, thereby altering the optical depth, as detailed in \citet{Perna2021a}. Additionally in a magnetized disk, accretion onto embedded stars may launch jets \citep[e.g.,][]{2005ApJ...629..397P,2024PASJ...76..823S}, which could clear out part of the accretion disk above each star and provide a low-optical-depth path along which photons can escape after stellar collapse. 

It is also important to note that, unlike a static transient—where the optical depth encountered by the radiation remains constant throughout the event (except for changes due to photoionization)—a GRB is produced by a relativistically expanding jet. The travel distance of this jet before it begins to decelerate may be comparable to the scale height of the AGN disk in some regions. Consequently, there will be areas in the disk where the early emission (produced when the jet head is still in the mid-plane and  $\tau>1$) is diffused, emerging on the diffusive timescale $t_{\rm diff}\sim H\tau/c$.
In contrast, the later, longer wavelength emission—produced when the jet’s emission front has propagated a considerable distance and encounters an optical depth $\tau<1$—may remain undiffused. For detailed quantitative analysis, refer to \citet{Wang2022}.

Therefore, the characteristics of GRB sources emerging from an AGN disk can differ significantly from those in standard galactic environments, exhibiting unique features. This distinction might make these sources potential probes of AGN disks themselves.
However, estimating an absolute observable rate is a rather challenging task, as the rate depends on the absolute number of stars in the disk (still poorly constrained), 
their radial distribution (addressed in this study), and the local opacity. A rough estimate for observable events per AGN can be made as follows: assuming a star formation rate $\rm SFR \sim 1 \, M_{\odot} \, \rm yr^{-1}$ and $\sim 1$ supernova per $100 \, M_{\odot}$ \citep[e.g.][]{2017ApJ...848...25M}, the SNe fraction is $f_{\rm SNe} \sim  0.01 M_{\odot}^{-1}$. The rate of observable events from the regions with canonical and massive star evolution can be parametrized using a factor for the optically thin, observable fraction of the disk containing evolving stars, $f_{\rm obs} = 0.1$ (fiducial value). This gives $\dot{N}_{\rm obs} = {\rm SFR} \cdot f_{\rm SNe} \cdot f_{\rm obs} \sim 10^{-3} \, \rm SNe \, yr^{-1}$ for a single AGN disk. More accurate estimates require the consideration of multiple additional factors \citep[e.g][]{2021MNRAS.507..156G,Kang2024}.

\begin{figure*} 
    \centering
    \includegraphics[width=1.\linewidth]{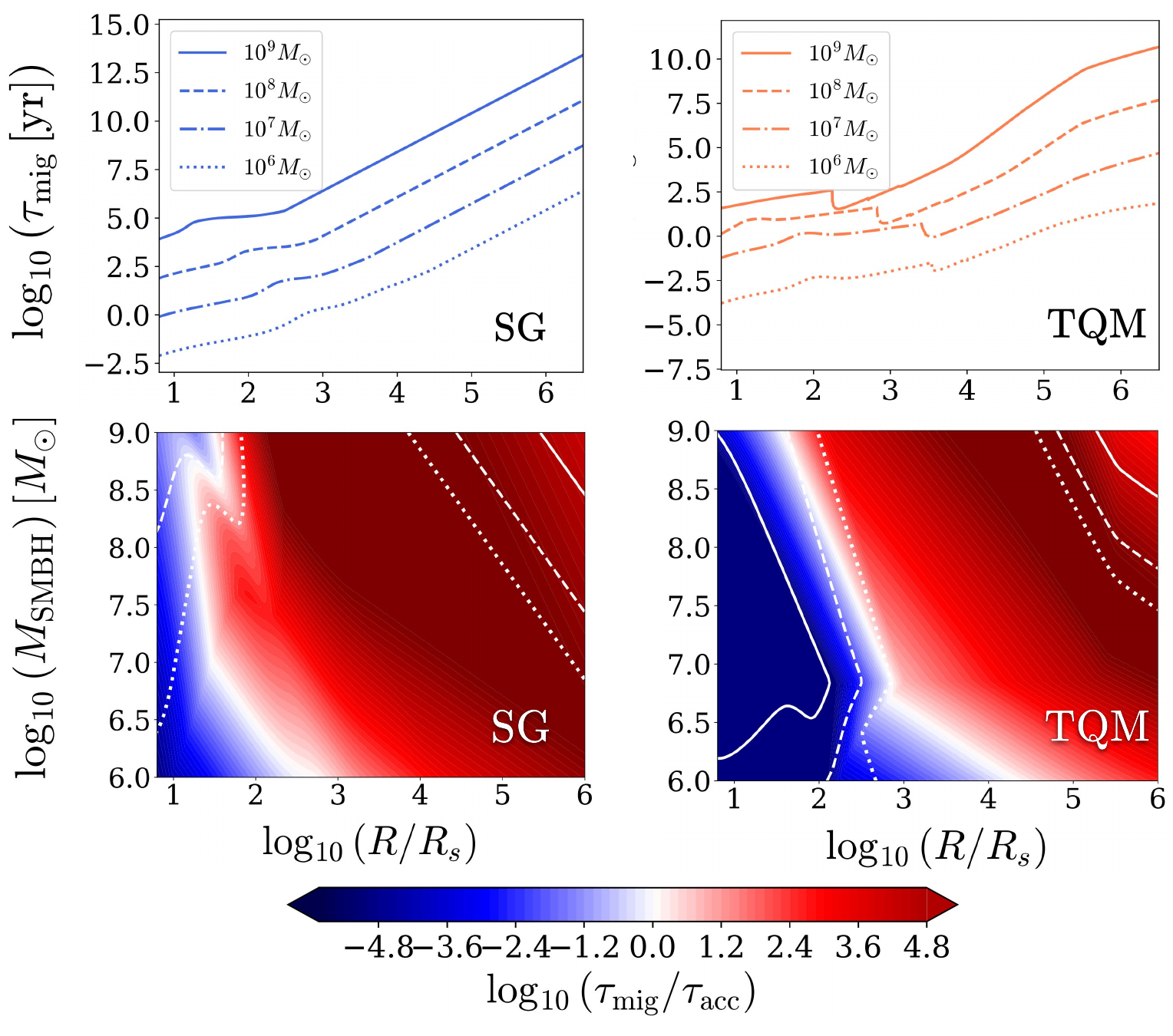}
    \caption{\textit{Upper panels}: Type I migration timescale $\tau_{\rm mig}$ (Eq.~\ref{eq:taumig}) based on the \citetalias{SG} and \citetalias{TQM} parameters for a range of SMBH masses ($10^6-10^9 M_{\odot}$). 
    \textit{Lower panels}: Contour map of the ratio between $\tau_{\rm mig}$ and $\tau_{\rm acc}$ (Eq.~\ref{eq:tau_M}) for the \citetalias{SG} and \citetalias{TQM} disk models, respectively. The blue region corresponds to where $\tau_{\rm mig}$ becomes shorter than $\tau_{\rm acc}$, and therefore where the evolutionary regions are likely to be affected by migration. Lower SMBH disks are more affected by the impact of migration. 
    } 
    \label{fig:mig}

\end{figure*}

\begin{figure*}
    \centering
    \includegraphics[width=1.\linewidth]{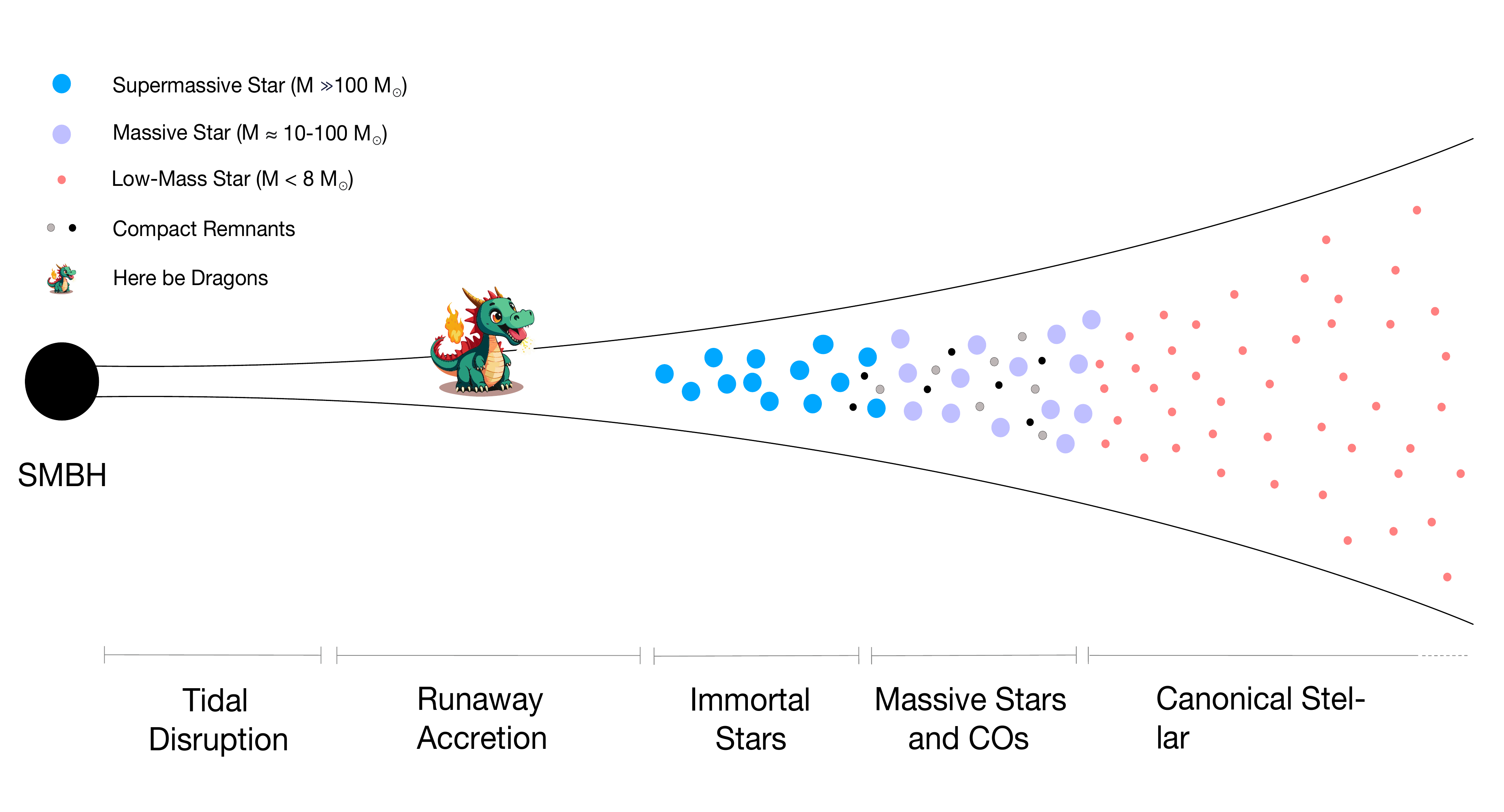}
    \caption{Illustration of different regimes of accretion in an AGN disk. Stellar seeds are expected to become massive and supermassive in the massive and immortal stars regions, respectively. While stars are expected to grow and become very massive in runaway accretion region as well, their exact fate is not yet fully understood. Massive stars pollute the disk with their super-Eddington winds and stellar explosions (SNe and GRBs). They also leave behind compact remnants that can dynamically interact and merge. In the slow accretion region, stars only gain modest amounts of material. These stars are expected to outlive the AGN phase. Note that the location of these different regions depends on the specific disk model, with this ordering  inspired by results for the SG model at low SMBH masses, see Fig.~\ref{fig:rhomap} and Fig.~\ref{fig:agn_chart}.}
    \label{fig:dragon}
\end{figure*}

\subsection{Canonical Stellar Evolution}

In regions of the disk where the accretion rate is low, stars are either not affected, or only marginally affected by accretion from the AGN disk. These stars evolve similarly to stars in typical ISM conditions.
We find this type of AGN stars in the  outer parts of the disk, at radii $\gtrsim$~a few $\times 10^5 R_{s}$. These outer zones are also regions where stars tend to form due to gravitational instabilities in the disk itself. In the \citetalias{TQM} model we also find canonical evolution very close to the SMBH, where the accretion rate is reduced substantially due to tidal effects combined with higher sound speeds and slightly lower gas densities in comparison to \citetalias{SG}. 
In these regions of canonical stellar evolution, the rate of massive stars experiencing a GRB accompanying a supernova is likely comparable to that in standard galactic environments. The ambient densities are also generally below the critical density $n_{\rm crit}$, which means the main features of typical GRB emission remain mostly unaltered.

The direct observability of these transients (both GRBs and supernovae) depends strongly on the disk model. The SG disk is fully transparent in the outermost regions, allowing emissions to be observed directly, while the TQM disk would cause the emission to emerge on the diffusive timescale. Therefore, observing these transients from an AGN disk may provide valuable insights into the disk's inner structure.

\newpage

\subsection{Possible Caveats due to Migration}

Throughout the paper we compute stellar accretion rates and timescales based on local disk properties, assuming that stars evolve at a fixed location in the disk. We acknowledge however, that the stellar migration is an important factor that can significantly alter the relative fraction of the stellar population in the various regions of the disk. For this reason, we do not estimate the number of stars contained in each evolutionary region. 
In order to estimate what kind of impact migration might have on the location of the different evolutionary bands, we analyze with a linear approach the migration rate $\tau_{\rm mig}$ experienced by stars due to Lindblad and linear corotation torques \citep{1979ApJ...233..857G,1997Icar..126..261W,2002ApJ...565.1257T}, roughly approximating

\begin{equation}\label{eq:taumig}
    \tau_{\rm mig} \approx \frac{GM_{\rm SMBH}}{\Omega^2R^2c_s}\left(\frac{M_*}{M_{\rm SMBH}}\right)^{-1} \left(\frac{H}{R}\right)^{2},
\end{equation}
where we use $M_*=1 \, M_{\odot}$. The upper panels of Fig.~\ref{fig:mig} show the migration timescale for \citetalias{SG} and \citetalias{TQM} respectively, for a range of $10^6-10^9 \, M_{\odot}$. These two panels show that as also inferred from Eq.~\ref{eq:taumig}, a heavier central black hole results in a longer $\tau_{\rm mig}$, while the trend of the curves resembles the dependence on the aspect ratio.

In order to estimate the rate of how fast a solar-type star moves through the disk compared to how fast it can accrete mass, we compare $\tau_{\rm mig}$ with the accretion timescale  from Eq.~\ref{eq:tau_M}. 
The lower panels of Fig.~\ref{fig:mig} show the contour map of the ratio between the two timescales ($\tau_{\rm mig}/\tau_{\rm acc}$). Over-plotted are the different evolutionary regions as in the previous sections. 
At $M_{\rm SMBH}\ge 10^7 \, M_{\odot}$ for both models (especially \citetalias{SG}), type I migration occurs 
at a slower rate than accretion for a large fraction of the parameter space, while this process becomes more important for the less massive central black holes, as migration is more efficient for smaller SMBHs.
Therefore, the location of the evolutionary regions in the outer parts of the disk does not seem to be affected by migration, while we expect the width of the bands in the inner regions to change, as the star does not accrete enough mass and moves through the disk experiencing different local gas densities and sound speeds.   
This analysis is a simple first step towards the understanding of a complex and non-linear problem, in which mass accretion 
changes while the star moves through the disk, as the migration timescale has also a mass dependence.
As previously stated in the paper, very massive stars could create gaps in the disk (requiring the implementation of type II migration) and alter the disk structure. 
We also do not take nonlinear and non-isothermal corotation
torques into account \citep[e.g.,][]{2010MNRAS.401.1950P, 2015ApJ...807L..11D}, or other effects that might stall or expedite migration \citep[e.g.][]{Bellovary16,2021PhRvD.103j3018P,2024MNRAS.530.2114G}.
Furthermore, gravitational radiation becomes important for distances $\lesssim 10^2 \, R_s$, which would accelerate migration in the inner regions.

\section{Summary}

The presence and evolution of stars and compact objects in AGN disks is currently a topic of great interest thanks to their potential contribution to LIGO/Virgo events, nuclear transients, and the effect of embedded objects on accretion disk structure.
It is therefore critical to assess the fraction of stars following specific evolutionary paths within the AGN disk. 
In this study, we have utilized the AGN disk models proposed by \citetalias{SG} and \citetalias{TQM} to analyze a broad range of SMBH masses, ranging from $10^6-10^9\rm{M}_\odot$. Our findings, summarized in Figure~\ref{fig:dragon}, 
schematically depict the key regions.  
Canonical stellar evolution dominates in both the outermost ad innermost regions of the disk, albeit for the \citetalias{SG} model this region is only confined to a very small band at the highest SMBH masses.
Moving towards the central regions of the disk, accretion from the surrounding medium becomes increasingly influential, although stars can still evolve until core collapse, potentially leading to transient events such as supernovae and GRBs. 

In regions of higher densities
and colder gas (particularly pronounced in the case of the \citetalias{TQM} model),
accretion plays a more dominant role. Eventually, a balance is reached where the accretion of fresh gas offsets the consumption of burnt fuel in the stellar core, allowing stars to persist indefinitely within the AGN disk. This is the so-called immortal branch of stars. This balance, however, is not realized in large swaths of the disk, where accretion takes over in a runaway fashion. The ultimate end state of this 'runaway population' is not well known, and coupled with the ability of the disk to continue feeding gas at the required high rate.
We have also explored the location of the different evolutionary regions in terms of the optical depth to map which kinds of stellar types reside in the optically thin regions and can therefore be "visible". We find that in the case of \citetalias{SG}, all the four stellar evolution types can reside in the optically thin region, while in \citetalias{TQM} models the disk is optically thick for the majority of the parameter space and only runaway stars reside in the visible region. 

Last, tidal effects near the SMBH prevent the survival of self-gravitating objects. The resulting regions without stars occupy a band which extends up to tens of 
Schwarzschild's 
radii for the smallest SMBH $\sim 10^6 M_\odot$, and becomes negligible above $\gtrsim 10^7 M_\odot$. 
The complex dynamics within AGN disks offer a rich tapestry of phenomena, highlighting the intricate interplay between the environment, stars, and compact objects.

\appendix

\section{Varying Disk parameters}\label{appendix:alpha}
\begin{figure*}[ht] 
    \centering
    \includegraphics[width=1.\linewidth]{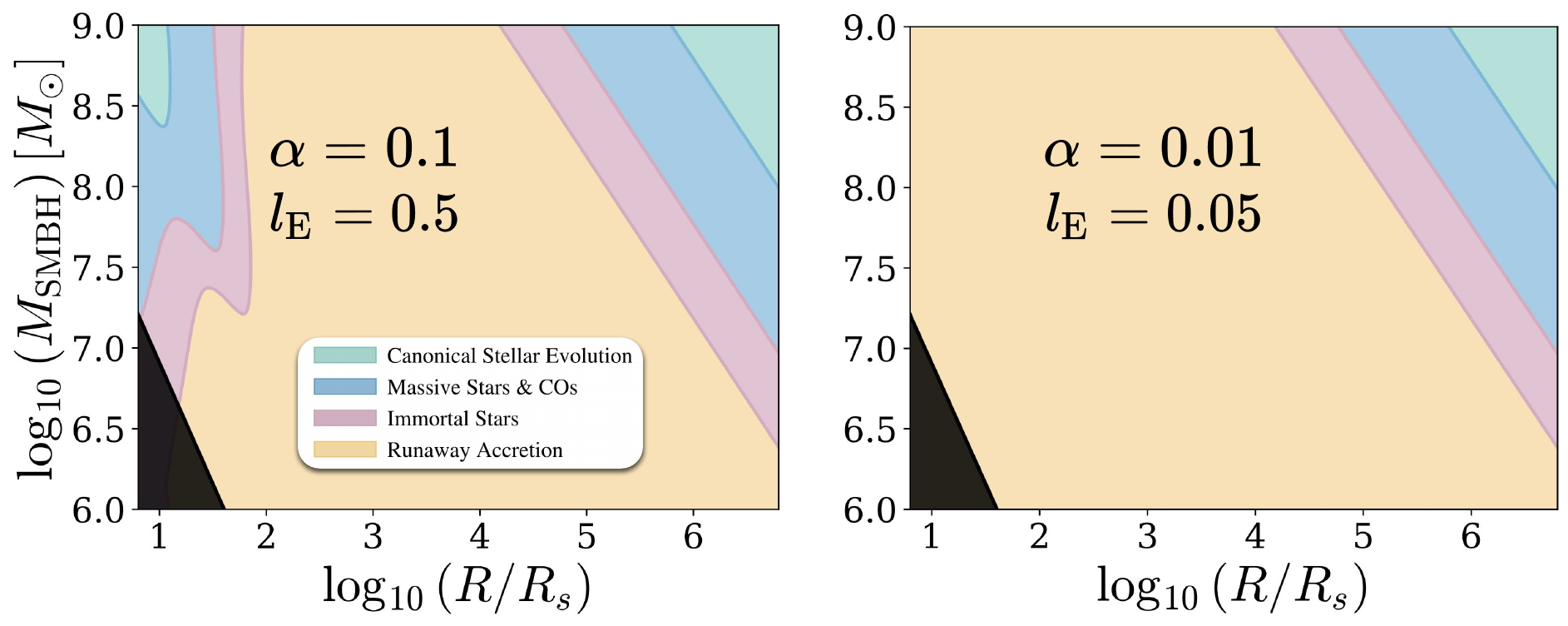}
    \caption{Different stellar evolution outcomes for the \citetalias{SG} disk model, for different values of the $\alpha$ viscosity and the Eddington ratio $l_{\rm E}$ of the disk.
    In the left panel we keep the original value of $l_{\rm E} =0.5$ but increase $\alpha$ to 0.1. In the right panel we lower $l_{\rm E}$ to 0.05 but keep the original value of $\alpha = 0.01$. Increasing $\alpha$ has the effect of decreasing the density in the inner regions, making the accretion timescale longer. Decreasing $l_{\rm E}$ results in lower sound speeds and higher densities in the inner regions, making stellar accretion very efficient. As a consequence, at lower $l_{\rm E}$ runaway accretion dominates for the entire SMBH mass range. 
    } 
    \label{fig:alphamap}

\end{figure*}

In this section we explore the impact of the choice of disk parameters on the location of the different evolutionary regions. In particular, we take the \citetalias{SG} model and investigate the effects of a higher viscosity $\alpha=0.1$ and a lower Eddington ratio $l_{\rm E}=0.05$. 
The left panel of Fig.~\ref{fig:alphamap} shows how the evolutionary regions vary increasing  $\alpha$ but keeping the original value of 0.5 for the Eddington ratio, while in the right panel we keep the original $\alpha$ and decrease $l_{\rm E}$. 
Increasing the viscosity results in a variation of the stellar population in the inner regions of the disk for $M_{\rm SMBH} \ge 10^7 \, \rm M_{\odot}$, where $\tau_{\rm acc}$ becomes longer, and we recover a small portion of canonical stellar evolution at high SMBH masses.  Stellar accretion becomes less efficient because in the \citetalias{SG} disk a higher $\alpha$ results in lower densities and temperatures in the inner regions, while at larger distances the stellar population remains unaffected, since in the outer regions the density does not depend on $\alpha$. 
On the other hand, decreasing $l_{\rm E}$ lowers the accretion rate, which results in a lower sound speed and higher densities. Consequently, stellar accretion is very efficient and $\tau_{\rm acc}$ becomes shorter than $\tau_{\rm KH}$, resulting in runaway accretion in the inner regions of the disk for the whole SMBH mass range.

\section*{Acknowledgements} 
The authors are grateful to Evgeni Grishin and the anonimous referee for useful comments and feedback on the paper. GF and JS are supported by Villum Fonden
grant No. 29466, and by the ERC Starting Grant no.
101043143 — BlackHoleMergs.
The Flatiron Institute is supported by the Simons Foundation.
AJD gratefully acknowledges support from LANL/LDRD under project number 20220087DR, and NASA ADAP grants 80NSSC21K0649 and 80NSSC20K0288. The LA-UR number is LA-UR-24-28457.
RP gratefully acknowledges support by NSF award AST-2006839.

\section*{Software and Data Availability}
We adopt the \citetalias{agnpack} Python package to obtain the accretion disk parameters for the \citetalias{SG} and \citetalias{TQM} models. Supporting data is available on Zenodo  \citep{zenodo}.

\bibliographystyle{mnras}
\bibliography{references}

\end{document}